# Statistical quality control for volumetric modulated arc therapy (VMAT) delivery using machine log data

Kwang-Ho Cheong, Me-Yeon Lee*, Sei-Kwon Kang, Jai-Woong Yoon, Soah Park,

Taejin Hwang, Haeyoung Kim, Kyoung Ju Kim, Tae Jin Han, Hoonsik Bae

*Department of Radiation Oncology, Hallym University College of Medicine, Anyang 431796, Korea*

The aim of this study is to set up statistical quality control for monitoring of volumetric modulated arc therapy (VMAT) delivery error using machine log data. Eclipse and Clinac iX linac with the RapidArc system (Varian Medical Systems, Palo Alto, USA) is used for delivery of the VMAT plan. During the delivery of the RapidArc fields, the machine determines the delivered motor units (MUs) and gantry angle position accuracy and the standard deviations of MU ($\sigma_{MU}$; dosimetric error) and gantry angle ($\sigma_{GA}$; geometric error) are displayed on the console monitor after completion of the RapidArc delivery. In the present study, first, the log data was analyzed to confirm its validity and usability; then, statistical process control (SPC) was applied to monitor the $\sigma_{MU}$ and $\sigma_{GA}$ in a timely manner for all RapidArc fields: a total of 195 arc fields for 99 patients. The $\sigma_{MU}$ and $\sigma_{GA}$ were determined twice for all fields, that is, first during the patient-specific plan QA and then again during the first treatment. The $\sigma_{MU}$ and $\sigma_{GA}$ time series were quite stable irrespective of the treatment site; however, the $\sigma_{GA}$ was strongly dependent on the gantry rotation speed. The $\sigma_{GA}$ of the RapidArc delivery for stereotactic body radiation therapy (SBRT) was smaller than that for the typical VMAT. Therefore, SPC was applied for SBRT cases and general cases respectively. Moreover, the accuracy of the potential meter of gantry rotation is important, since the $\sigma_{GA}$ can be dramatically changed due to its condition. By applying SPC to the $\sigma_{MU}$ and $\sigma_{GA}$, we could monitor the delivery error efficiently. However, the upper and lower limits of SPC need to be determined carefully with full knowledge of the machine and log data.

PACS number: 87.53.Tf, 87.56.-v, 87.53.St, 87.53.Xd

Keywords: IMRT, VMAT, RapidArc, SBRT, SPC

Email: mylee@hallym.or.kr

Fax: +82-31-380-3913

# I. INTRODUCTION

The goal of quality control (QC) or quality assurance (QA) is to quantify systematic errors in order to maintain the quality of a certain process or equipment. In the traditional approaches to QA, a measurement is performed and then compared with a predetermined specification. If the specification is met, no further action is taken, but if not, an additional investigation is implemented to determine the causes and to apply correction. Thus, only a binary decision is made concerning the measurement. In this case, the tolerance limit is an important issue [1,2]. QA guidelines such as AAPM TG 142 are based on this concept [3]. Modern QA tools, by contrast, generally deal with time-ordered data, metrics or data from process performance to draw inductive conclusions [1,4]. Understanding how a process is performing, then, is important. Moreover, data-variation trends also are monitored.

Statistical process control (SPC) has been utilized for radiation therapy QC since 2000 [4-9]. Its purpose is to maintain a process within the given specifications. SPC was first suggested in 1924 by Walter Shewhart, who developed the control chart and the concept of a state of statistical control, both of which are used to evaluate the stability of a process. If a variability of a predictable process is greater than the relevant specifications, the process is unacceptable and one should find a cause that affects the process.

In the medical field, volumetric modulated arc therapy (VMAT), now widely employed, has overtaken conventional intensity-modulated radiation therapy (IMRT) [10,11]. The VMAT process is very complex. During beam-on time, the dose rate changes, the gantry rotates, and the multi-leaf collimator (MLC) leaves also move. As such, treatment can be completed within only one or two gantry rotations. However, for realistic reasons, there are many assumptions made with respect to the process. For example, typical radiation therapy planning systems cannot exactly mimic the actual motion of the machine. Instead, an arc is approximated by some static beam and multiple segments, which are controlled as control points (CPs). CPs are not a new concept but have long been used in the DICOM RT standard. For VMAT, CPs roughly represent the beam delivery parameters at certain gantry angles. For example, in RapidArc (Varian Medical Systems, Palo Alto, USA), the number of CPs per arc is fixed at 178. VMAT delivery is determined by the delivery control system of the linac. The CPs are used to control the treatment in this procedure. For the Varian C-series linac, the Clinac controller controls the monitor units (MUs) versus gantry rotation, while the MLC controls MLC motion versus gantry rotation. During treatment, linac confirms the delivered MUs and gantry angle position, and then displays the results on the monitor approximately every 50 ms. A brief diagram of VMAT planning to delivery and verification for C-series linac is presented in Figure 1. In the case of TrueBeam (Varian Medical Systems, Palo Alto, USA), the so-called supervisor controls all nodes and motion axes. Therefore, the most important VMAT issue is how treatment is accurately delivered as planned.

The aim of this study is to set up a statistical quality control for monitoring of VMAT delivery error trends using machine log data for the RapidArc system. The variables dealt with were the standard deviations (SDs) of MUs ($\sigma_{MU}$) and gantry rotation position ($\sigma_{GA}$) after VMAT delivery. The results derived in this study suggest a simple means of confirming and monitoring delivery error during treatment sessions.

## II. MATERIALS AND METHODS

### 1. Meaning of standard deviations in delivery result

Prior to applying SPC to the SDs of MUs and gantry angle rotation, their physical meaning in the delivery result was validated to determine their feasibility as monitoring variables. For this purpose, we compared the log data with the plan data to confirm their validity and usability. Eclipse ver. 10 (Varian Medical Systems, Palo Alto, USA) and Clinac iX with RapidArc system (Varian Medical Systems, Palo Alto, USA) were used for VMAT planning and delivery, respectively.

Figure 2 shows the two ways in which the SDs of the MUs and gantry rotation are confirmed. They are shown on the monitor just after completion of VMAT delivery, or alternatively, they can be seen in the dynalog data. However, as the data are only temporally displayed on the monitor, they should be recorded promptly, before they disappear. Dynalog data contain a summary of the delivery and statistical parameters, as shown in Figure 2(b); they also record the planned and actually delivered dose and gantry angle, as indicated in Figures 2(c) and (d). Figure 3 shows an example of a prostate case. One arc was used, and the MUs numbered 446. The number of CPs was 178, and the number of samplings was almost 1500. The recorded $\sigma_{MU}$ and $\sigma_{GA}$ were 0.06 and 0.17, respectively. Table 1 shows the results of the calculation of the mean differences between the planned and delivered MUs and gantry rotation position values. The converted data represents the cumulative MUs at each CP as well as the gantry angles converted to Varian scale from the raw Eclipse data. The planned and actual data were obtained from the dynalog data. First, the converted and planned data were compared to determine whether the planned data were correctly applied. In each of the rows, the MU and GA differences also are respectively calculated. However, the calculated mean differences were smaller than the offered values of 0.06 and 0.17, seemingly due to the sampling method: linac considers the overall error at every sampling point. In either case, $\sigma_{MU}$ and $\sigma_{GA}$ are not actual standard deviations in the statistical sense, but just the mean differences among the samples. However, hereafter, to avoid confusion, we will use the term "standard deviation."

**2. Statistical analysis of dosimetric and geometric standard deviations using SPC**

In this study, a total of 195 arc fields for 99 patients treated in our facility between January 2013 and September 2014 were used in the analysis, as shown in Table 2. The prostate cancer patients were the most numerous, followed by the head & neck cancer patients; however, there were more plans and arcs for the head & neck cancer patients. The typical VMAT (hereafter: IMRT) arcs numbered 153 of 195 (78.5%), while the VMAT arcs for stereotactic body radiation therapy (SBRT) numbered 42 of 195 (21.5%).

SPC was applied to an individual arc as a data point, because the $\sigma_{MU}$ and $\sigma_{GA}$ were recorded for each arc. The $\sigma_{MU}$ and $\sigma_{GA}$ were determined twice for all fields: first during the patient-specific plan QA, and then during the first treatment. However, discrepancy between two observations was relatively small. In other words, considering ranges(R) was meaningless in this case. Therefore, we employed the individuals and moving range (I-MR) chart as the SPC process behavior charts (PBCs) to detect large changes in the process [2]. The I-MR chart consists of an upper control limit (UCL), a center line (CL), and a lower control limit (LCL), as shown in Eqs. (1) to (3),

$$\text{UCL} = \bar{X} + 3\frac{\overline{MR}}{d_2} \qquad (1)$$

$$\text{CL} = \bar{X} \qquad (2)$$

$$\text{LCL} = \bar{X} - 3\frac{\overline{MR}}{d_2} \qquad (3),$$

where CL is the mean of individual points, $\overline{MR}$ is the mean of differences between consecutive observations ($MR_i = |X_i - X_{i-1}|$) and $d_2$ is a divisor for estimation of variance for the process. We used $d_2$=1.128 because subgroup size(n) was 2 in this study. Moreover, we set the width of control to 6σ, which is typically used in the field.

We also used the process capability index ($C_p$) and process acceptability index ($C_{pk}$) to investigate the variation process of the data with respect to the upper and lower action limits (UAL and LAL) and the allowed width of control [12]. The greater the $C_p$ value, the better able is the process to meet the action limits (Eq. (4)). $C_{pk}$ indicates how close the process center is to the nearest action limit, and is calculated from Eq. (5). When calculating $C_p$ and $C_{pk}$, it is important that the process is under control, that is, that no points are outside the control limits on the $\bar{X}$-chart.

$$C_p = \frac{UAL - LAL}{6 \cdot \sigma} \qquad (4)$$

$$C_{pk} = \min\left(\frac{UAL - \mu}{3 \cdot \sigma}, \frac{\mu - LAL}{3 \cdot \sigma}\right) \qquad (5)$$

We used Minitab ver. 16 (Minitab Inc, State College, USA), a QC-specialized statistical package, for the statistical analysis.

### III. RESULTS

Figure 4 shows the overall variations of $\sigma_{MU}$ and $\sigma_{GA}$. The time series were quite stable regardless of the treatment site. The many valleys in the graph are due to the SBRT cases, in which the gantry rotation speed was slower than in IMRT, and the $\sigma_{GA}$, correspondingly, was relatively smaller. Therefore, it is necessary to separate SBRT cases from routine IMRT ones. Also in the graph, there is a step change that possibly is due to the replacement of the potential meter for the gantry rotational position on March, 2014. Accordingly, the general $\sigma_{GA}$ was decreased from 0.4 to 0.2 and stabilized.

Figures 5 and 6 are respective I-MR charts for all $\sigma_{MU}$ and $\sigma_{GA}$. The dosimetric error was well controlled through the process, even though there were some escaped values from the UCL or LCL. The LCL was meaningless here; thus, values over the UCL needed to be carefully monitored. Moving range was maintained less than 0.02 for most points as well. However, as shown in Figure 6, there were extreme variations and a step change. Therefore, we divided the overall $\sigma_{GA}$ data set into 4 groups, IMRT1 and IMRT2, and SBRT1 and SBRT2, the numbers 1 and 2 implying before and after the replacement of a potential meter, respectively.

Figure 7 provides the I-MR charts for the $\sigma_{GA}$ of each group. A center line shift is observable when we compare Figures 7(a) with (b) and (c) with (d). However, the widths of control levels (UCL-LCL) for each group are not varied much, or even increased due to variance inside the group. Figure 8 plots the capability histograms and provides the $\sigma_{GA}$ distributions for each group. The lower and upper specification levels (LSL and USL) were set to 0 and 0.5, respectively, in all of the cases. Each normal distribution before and after the potential meter replacement is obviously shifted to the lower error direction (LSL=0); thus, each can be considered to be stabilized. In both figures, points those exceed the control lines are presented in red color; only one is observed in SBRT1. Therefore, we can confirm that dosimetric and geometric error were acceptable until now.

Table 3 provides a summary of the analysis results. The UCL, CL, LCL, UCL-LCL, and $C_p$, $C_{pk}$ of the $\sigma_{MU}$ for all of the groups were similar, while the $\sigma_{GA}$ parameters had been more changed. A larger $C_p$ is indicative of good process capability and acceptability. For the IMRT and SBRT cases, it is obvious that the $C_{pk}$ was improved after replacement of the potential meter. Therefore, when SPC and process capability and acceptability are employed, one should review all of the data carefully before drawing any conclusions.

## IV. DISCUSSION

This was a retrospective study that considered all of the available data of already-treated patients. Thus, the mean value ($\bar{X}$) was estimated from all data points. Even so, for prospective purposes, control limits needed to be set based on certain data points, generally first some data. To this end, the use of at least 20 data points is recommended; however, if there are no severe variation changes, around 10 data points usually is sufficient [12].

SPC is a versatile tool for at-a-glance monitoring of error trends over a given time period. It has been used to confirm the linac out constancy [12] and for patient-specific IMRT and VMAT QA [5,6,13-15]. Beam flattening or symmetry also can be monitored using SPC [16]. Simply, SPC can be applied to all QA aspects of radiotherapy, and potentially can substitute for the conventional QA paradigms. However, there are a number of concerns about the application of SPC to QC/QA procedures. First, a user should know which SPC analysis is most appropriate for the intended QC/QA purpose. In other words, SPC method selection is based on the data type or goal of a QC/QA procedure. If long-term data are used for analysis or if gradual drift is a main concern, then an exponentially weighted moving average (EWMA) chart might be more helpful [12]. Thus, for medical physicists, a full understanding of SPC and the ability to interpret the results and catch abnormalities within a data set also are required. When a data point exceeds an action threshold, a physicist should find the causes and correct them.

The other issues regarding SPC are related to data size: the total number of data points as well as subgroup size [8,9]. Subgroup size affects the calculation of the process behavior limits. In this study, we used 2 as a subgroup size, which means that the data was collected two times in a short time period: first during the patient-specific QA time, and then again at the first treatment time, as already noted. Subgroup size is especially important in R-chart when determining range variations. However, in the case of VMAT delivery QA, the data in a subgroup is almost identical; thus it seems that I-MR chart with single-measurement data is sufficient for dosimetric and geometric error monitoring. The $\sigma_{MU}$ and $\sigma_{GA}$ values obtained in this study are pertinent only to our facility, as value ranges can vary from hospital to hospital. The gantry rotation position is especially strongly dependent on the condition of the potential meter, as emphasized earlier. Therefore, careful investigation of $\sigma_{GA}$ variation should be conducted after any potential meter replacement or recalibration. As we observed in the results section above, the gantry rotation speed affects the $\sigma_{GA}$: if the gantry rotates slowly, the $\sigma_{GA}$ will be smaller than in typical VMAT delivery. This happens when the MUs at certain CP exceeds the deliverable MU limits with the maximum dose rate (600 MU/min for Varian C-series linac). According to the present results, the gantry rotation direction slightly affected the $\sigma_{GA}$, but not to the point of statistical significance. And neither is the $\sigma_{GA}$ significantly influenced by treatment sites such as the head & neck, prostate, chest, or pelvic region.

We did not verify the values in the log data by independent measurement, because the trend of variation is more important than the accuracy of the individual value of each data point.

## V. CONCLUSIONS

We were able to monitor dosimetric and geometric VMAT delivery errors simply but efficiently by application of SPC. This notwithstanding, the upper and lower limits of SPC need to be determined carefully with full knowledge of the machine and log data. Moreover, the analysis of SDs of IMRT and SBRT needs to be separated, especially with respect to the $\sigma_{GA}$. Moreover, the $\sigma_{GA}$ is strongly dependent on the condition of the potential meter of gantry rotation.

## ACKNOWLEDGEMENT

This work was supported by the Radiation Technology R&D program through the National Research Foundation of Korea funded by the Ministry of Science, ICT & Future Planning (no. 2013043498).

**REFERENCES**

[1] T. Pawlicki, and A.J. Mundt, Med.Phys. **34,** 1529 (2007).

[2] T. Sanghangthum, S. Suriyapee, S. Srisatit, and T. Pawlicki, J.Radiat.Res. **54,** 546 (2013).

[3] E.E. Klein, J. Hanley, J. Bayouth, F.F. Yin, W. Simon, S. Dresser, C. Serago, F. Aguirre, L. Ma, B. Arjomandy, C. Liu, C. Sandin, T. Holmes, and Task Group 142, American Association of Physicists in Medicine, Med.Phys. **36,** 4197 (2009).

[4] T. Pawlicki, B. Chera, T. Ning, and L.B. Marks, Semin.Radiat.Oncol. **22,** 70 (2012).

[5] K. Gerard, J.P. Grandhaye, V. Marchesi, H. Kafrouni, F. Husson, and P. Aletti, Med.Phys. **36,** 1275 (2009).

[6] G. Palaniswaamy, R. Scott Brame, S. Yaddanapudi, D. Rangaraj, and S. Mutic, Med.Phys. **39,** 7560 (2012).

[7] S. Qin, M. Zhang, S. Kim, T. Chen, L.H. Kim, B.G. Haffty, and N.J. Yue, Radiat.Oncol. **8,** 225 (2013).

[8] T. Pawlicki, M. Whitaker, and A.L. Boyer, Med.Phys. **32,** 2777 (2005).

[9] T. Pawlicki, and M. Whitaker, Int.J.Radiat.Oncol.Biol.Phys. **71,** S210 (2008).

[10] C.X. Yu, and G. Tang, Phys.Med.Biol. **56,** R31 (2011).

[11] K. Otto, Med.Phys. **35,** 310 (2008).

[12] T. Sanghangthum, S. Suriyapee, S. Srisatit, and T. Pawlicki, J.Appl.Clin.Med.Phys. **14,** 4032 (2013).

[13] S.L. Breen, D.J. Moseley, B. Zhang, and M.B. Sharpe, Med.Phys. **35,** 4417 (2008).

[14] J.D. Gagneur, and G.A. Ezzell, J.Appl.Clin.Med.Phys. **15,** 4927 (2014).

[15] G.M. Mancuso, J.D. Fontenot, J.P. Gibbons, and B.C. Parker, Med.Phys. **39,** 4378 (2012).

[16] C.M. Able, C.J. Hampton, A.H. Baydush, and M.T. Munley, Radiat.Oncol. **6,** 180 (2011).

Table 1. Calculation results of mean differences between planned and delivered values of MUs and gantry rotation position. The converted data represents the cumulative MUs at each control point and gantry angles converted to the Varian scale from the raw Eclipse data. The planned and actual data came from the dynalog data.

| CP # | From Eclipse | | | | From Dynalog | | | | MU$_{diff}$ | GA$_{diff}$ |
|---|---|---|---|---|---|---|---|---|---|---|
| | raw data | | converted data | | planned data | | actual data | | | |
| | weight | GA | MU | GA | MU | GA | MU | GA | | |
| 1 | 0.00 | 181.00 | 0.00 | 359.00 | 0.00 | 359.00 | 0.00 | 359.03 | **0.00** | **0.03** |
| 2 | 0.00 | 182.00 | 1.34 | 358.00 | 1.33 | 358.00 | 1.38 | 358.45 | **0.05** | **0.45** |
| 3 | 0.01 | 184.10 | 3.97 | 355.90 | 3.98 | 355.90 | 4.04 | 355.95 | **0.06** | **0.05** |
| 4 | 0.01 | 186.10 | 6.65 | 353.90 | 6.64 | 353.90 | 6.67 | 353.98 | **0.03** | **0.08** |
| 5 | 0.02 | 188.10 | 9.32 | 351.90 | 9.30 | 351.90 | 9.34 | 351.88 | **0.04** | **0.02** |
| 6 | 0.03 | 190.20 | 12.00 | 349.80 | 11.99 | 349.80 | 12.04 | 349.90 | **0.05** | **0.10** |
| 7 | 0.03 | 192.20 | 14.72 | 347.80 | 14.71 | 347.80 | 14.75 | 347.95 | **0.04** | **0.15** |
| 8 | 0.04 | 194.20 | 17.44 | 345.80 | 17.44 | 345.80 | 17.48 | 345.98 | **0.04** | **0.18** |
| 9 | 0.05 | 196.30 | 20.16 | 343.70 | 20.16 | 343.70 | 20.19 | 343.90 | **0.03** | **0.20** |
| 10 | 0.05 | 198.30 | 22.84 | 341.70 | 22.83 | 341.70 | 22.87 | 341.85 | **0.04** | **0.15** |
| | | | | …… | | | | | | |
| 171 | 0.96 | 165.80 | 429.23 | 14.20 | 429.22 | 14.20 | 429.27 | 14.40 | **0.05** | **0.20** |
| 172 | 0.97 | 167.80 | 431.82 | 12.20 | 431.80 | 12.20 | 431.83 | 12.45 | **0.03** | **0.25** |
| 173 | 0.97 | 169.80 | 434.36 | 10.20 | 434.38 | 10.20 | 434.42 | 10.48 | **0.04** | **0.28** |
| 174 | 0.98 | 171.90 | 436.95 | 8.10 | 436.96 | 8.10 | 437.01 | 8.38 | **0.05** | **0.28** |
| 175 | 0.99 | 173.90 | 439.53 | 6.10 | 439.54 | 6.10 | 439.57 | 6.28 | **0.03** | **0.18** |
| 176 | 0.99 | 175.90 | 442.12 | 4.10 | 442.12 | 4.10 | 442.16 | 4.38 | **0.04** | **0.28** |
| 177 | 1.00 | 178.00 | 444.71 | 2.00 | 444.71 | 2.00 | 444.75 | 2.25 | **0.04** | **0.25** |
| 178 | 1.00 | 179.00 | 446.00 | 1.00 | 446.00 | 1.00 | 446.05 | 1.05 | **0.05** | **0.05** |
| **Mean differences** | | | | | | | | | **0.04** | **0.15** |

* Abbreviations: GA: gantry angle, MUs: monitor units

Table 2. Number of patients, plans and arcs per treatment site. A total of 195 arc fields for 99 patients treated in our facility from January 2013 to September 2014 were used in the analysis.

|  | Number of patients | Number of plans | Number of arcs |
| --- | --- | --- | --- |
| Head & neck | 25 | 51 | 77 |
| Prostate | 38 | 43 | 49 |
| Lung SBRT | 18 | 19 | 37 |
| Brain | 9 | 10 | 14 |
| Chest | 3 | 5 | 5 |
| Bone SBRT | 3 | 3 | 5 |
| Cx, vagina | 2 | 3 | 4 |
| Anal | 1 | 2 | 4 |
| Sum | 99 | 136 | 195 |

* Abbreviations: SBRT: stereotactic body radiation therapy

Table 3. Summary of analysis results. $C_p$ is a process capability index, and $C_{pk}$ is a process acceptability index. $\sigma_{MU}$ and $\sigma_{GA}$ represent the standard deviations of MUs and gantry angle positions, respectively.

|  |  | UCL | CL | LCL | UCL-LCL | $C_p$ | $C_{pk}$ |
|---|---|---|---|---|---|---|---|
| $\sigma_{MU}$ | all | 0.075 | 0.053 | 0.03 | 0.045 | 11.05 | 2.33 |
|  | IMRT1 | 0.074 | 0.055 | 0.036 | 0.038 | 13.04 | 2.85 |
|  | IMRT2 | 0.068 | 0.05 | 0.032 | 0.036 | 13.89 | 2.78 |
|  | SBRT1 | 0.070 | 0.048 | 0.025 | 0.045 | 11.06 | 2.11 |
|  | SBRT2 | 0.105 | 0.058 | 0.012 | 0.093 | 5.37 | 1.25 |
| $\sigma_{GA}$ | all | 0.417 | 0.273 | 0.130 | 0.287 | 1.74 | 1.58 |
|  | IMRT1 | 0.449 | 0.384 | 0.319 | 0.13 | 3.85 | 1.79 |
|  | IMRT2 | 0.306 | 0.220 | 0.134 | 0.172 | 2.91 | 3.26 |
|  | SBRT1 | 0.283 | 0.176 | 0.068 | 0.215 | 2.32 | 1.63 |
|  | SBRT2 | 0.116 | 0.08 | 0.044 | 0.072 | 6.96 | 2.23 |

* Abbreviations: UCL: upper control limit, CL: center line, LCL: lower control limit

**Figures**

Figure 1. Brief diagram of VMAT planning to delivery and verification for C-series linac. Dynalog data for linac and MLC are displayed separately on the computer monitor.

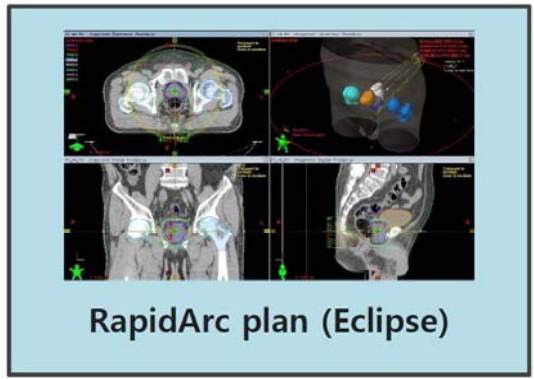
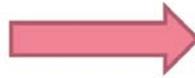
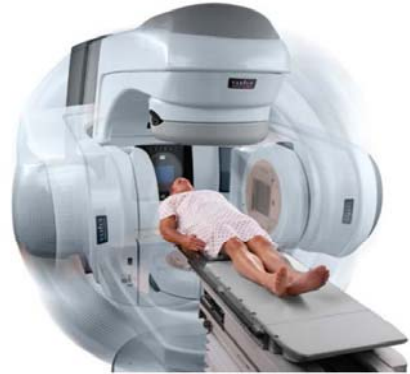
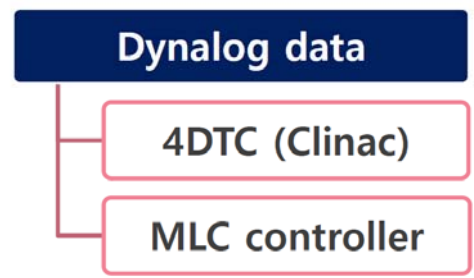

Figure 2. Two ways to confirm standard deviations of MUs and gantry rotation: (a) on the console monitor just after completion of VMAT delivery or (b-d) in the dynalog data

(a)

(b)

```
                -- STT --
INSTANCE#    DOSE      GANTRY ANGLE    (Internal scale)
             (MU)         (deg)
    1        0.00        359.00
    2        1.60        358.00
    3        4.80        355.90
    4        8.00        353.90
    5       11.20        351.90
    6       14.40        349.80
    7       17.60        347.80
    8       20.81        345.80
    9       24.00        343.70
   10       27.16        341.70
   11       30.29        339.70
   12       33.41        337.60
   13       36.54        335.60
   14       39.66        333.60
   15       42.78        331.50
   16       45.90        329.50
   17       49.02        327.50
   18       52.12        325.40
   19       55.20        323.40
   20       58.28        321.40
   21       61.36        319.30
-- More --_
```

(c)

```
  166        520.49       24.40
  167        523.82       22.40
  168        527.15       20.30
  169        530.49       18.30
  170        533.89       16.30
  171        537.37       14.20
  172        540.86       12.20
  173        544.34       10.20
  174        547.81        8.10
  175        551.29        6.10
  176        554.78        4.10
  177        558.26        2.00
  178        560.00        1.00

   -- SEGMENT BOUNDARY SAMPLES (ACTUAL) --
INSTANCE#    DOSE      GANTRY ANGLE    (Internal scale)
             (MU)         (deg)
    1        0.00        359.00
    2        1.65        358.35
    3        4.84        356.33
    4        8.05        354.28
    5       11.24        352.20
    6       14.43        350.13
    7       17.65        348.08
-- More --_
```

(d)

Figure 3. Example of prostate case. One arc was used, and the MUs numbered 446. The number of CPs was 178, and the number of samplings was almost 1500. The recorded $\sigma_{MU}$ and $\sigma_{GA}$ were 0.06 and 0.17, respectively.

```
CLINAC IX-S - SN 992

DATE..................: 04/28/2014
TIME..................: 21:10:09

            D Y N A M I C   B E A M   D E L I V E R Y   L O G   F I L E

               ** T R E A T M E N T   S E T U P **

TREATMENT TYPE : VMAT
ENERGY         : 10 X
MU             : 446
TIME           : 3.00 (min)
ACCESSORY      : NO ACCESSORY

               ** D Y N A M I C   B E A M   S T A T I S T I C S **

TOTAL DOSE DELIVERED              :   446 (MU)
DOSE STANDARD DEVIATION           :   0.06 (MU)
DOSE-POSITION STANDARD DEVIATION  :   0.17 (deg)
NUMBER OF SAMPLES                 :   1510

-- More --
```

Figure 4. Overall variations of $\sigma_{MU}$ and $\sigma_{GA}$. Time series of $\sigma_{MU}$ and $\sigma_{GA}$ were quite stable regardless of the treatment site. There were many valleys in the graph, due to SBRT cases wherein the gantry rotation speed was slower than in IMRT ones, and thus the $\sigma_{GA}$ was relatively smaller than in IMRT. Also there is a step change that possibly is due to the replacement of the potential meter for the gantry rotational position on March, 2014. Accordingly, the general $\sigma_{GA}$ was decreased from 0.4 to 0.2 and stabilized.

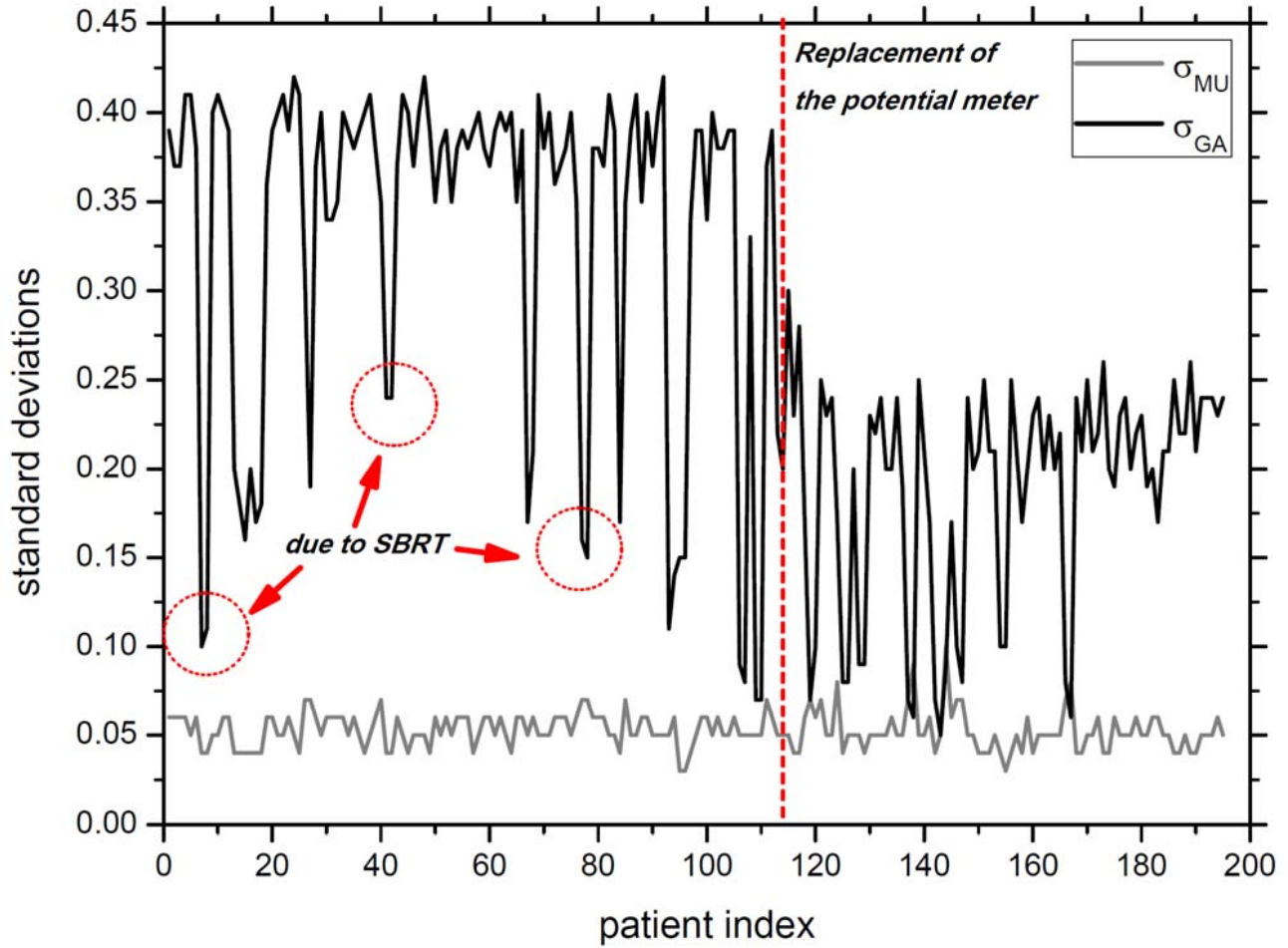

Figure 5. The individual observations (I chart) and moving range (MR chart) for all $\sigma_{MU}$. The center line (CL) is presented in a green line and the upper and lower control limits (UCL and LCL each) are shown in red lines. The dosimetric error is well controlled through the process, even though there are some escaped values from the UCL or LCL (red square symbols).

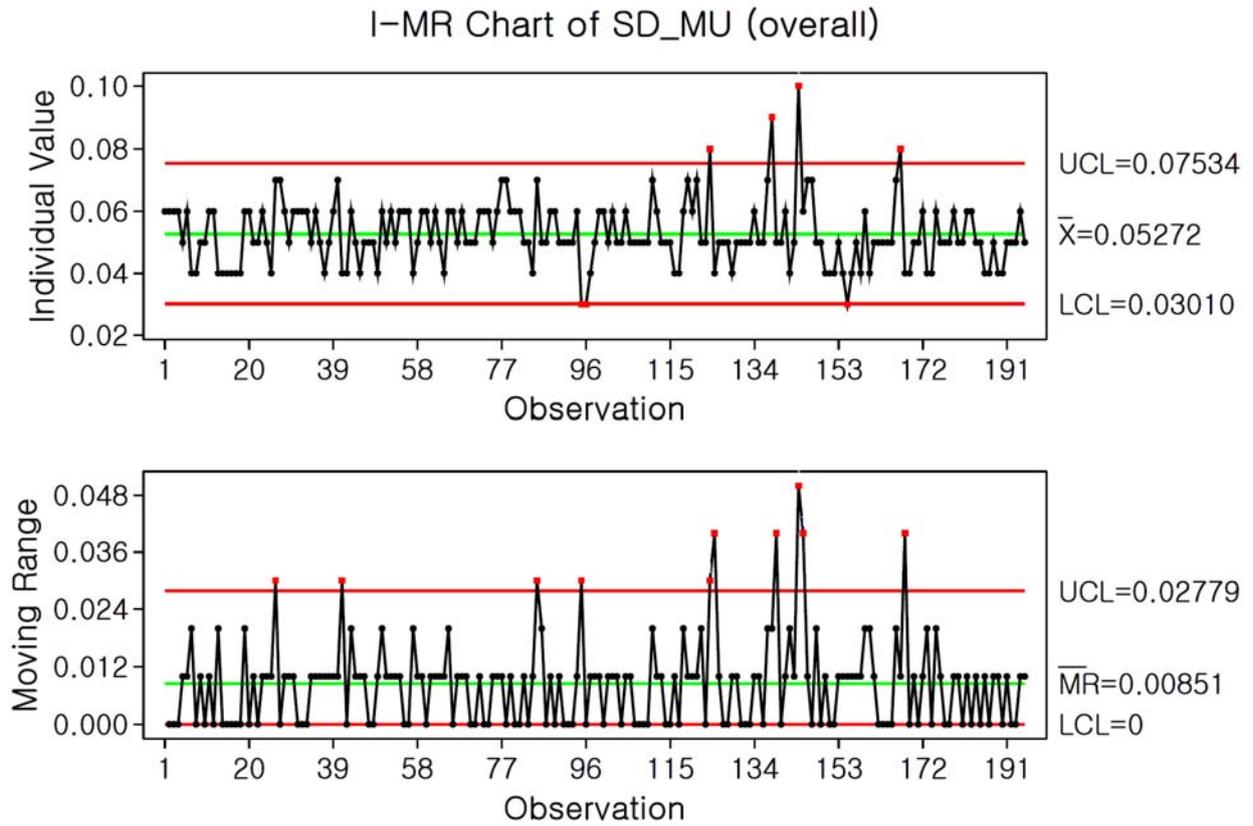

Figure 6. The individual observations (I chart) and moving range (MR chart) for all $\sigma_{GA}$. The center line (CL) is presented in a green line and the upper and lower control limits (UCL and LCL each) are shown in red lines. The variation of geometric error is relatively larger than that of dosimetric error, and a step change presents. However, there are very few escaped values from the UCL.

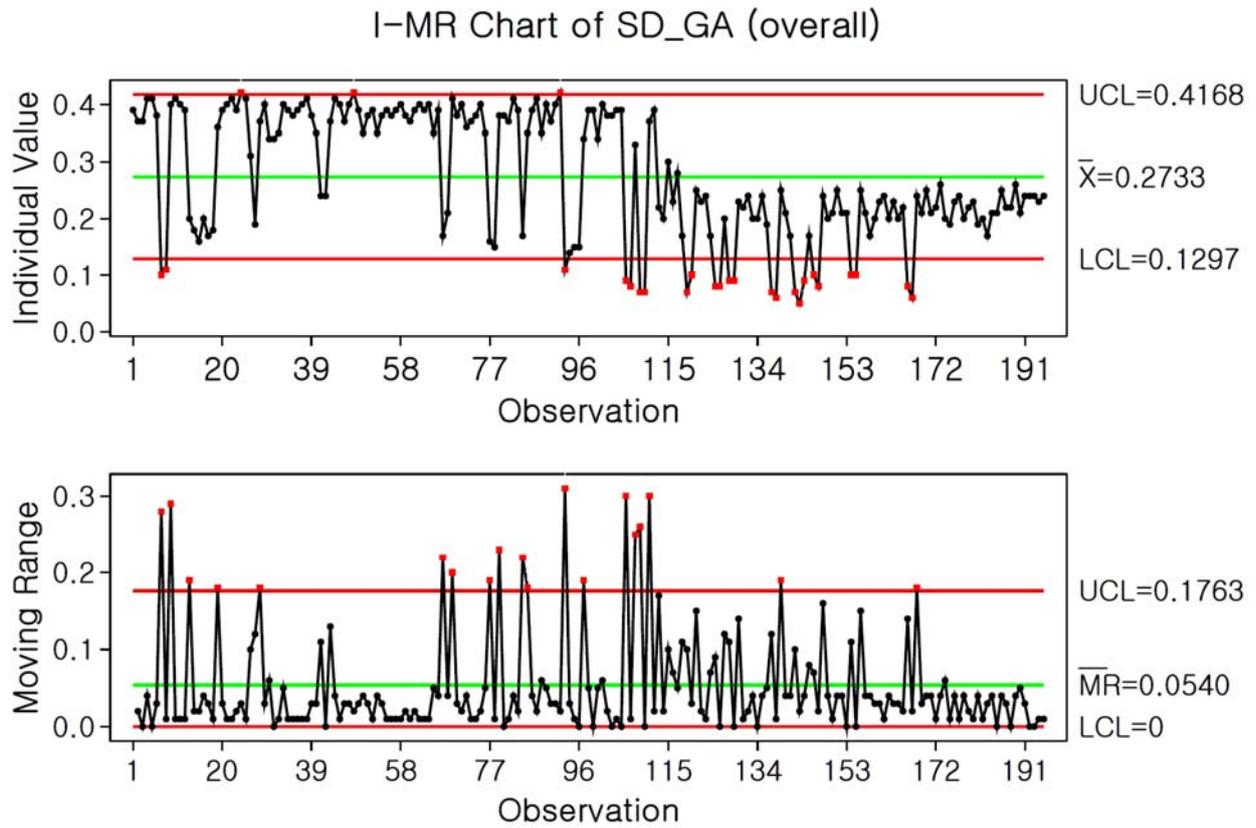

Figure 7. I-MR charts for σ_GA of each group. A center line shift is observed when we compare Figure 7(a) with (b) and (c) with (d). The UCL and LCL width of IMRT 2 became narrower than those of IMRT1; however, the control level width of σ_GA for SBRT was similar.

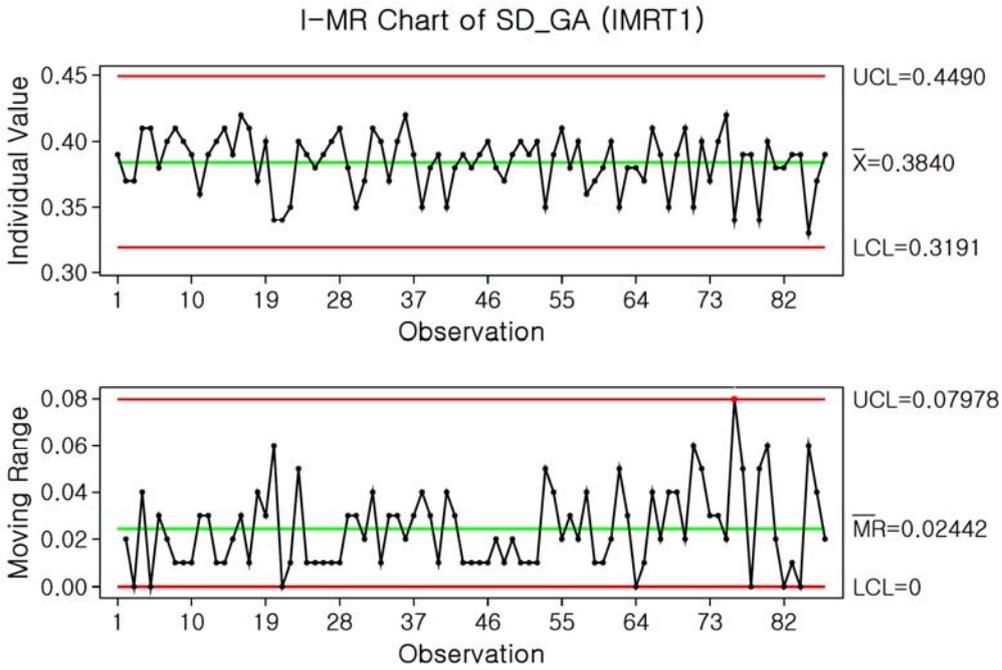

(a)

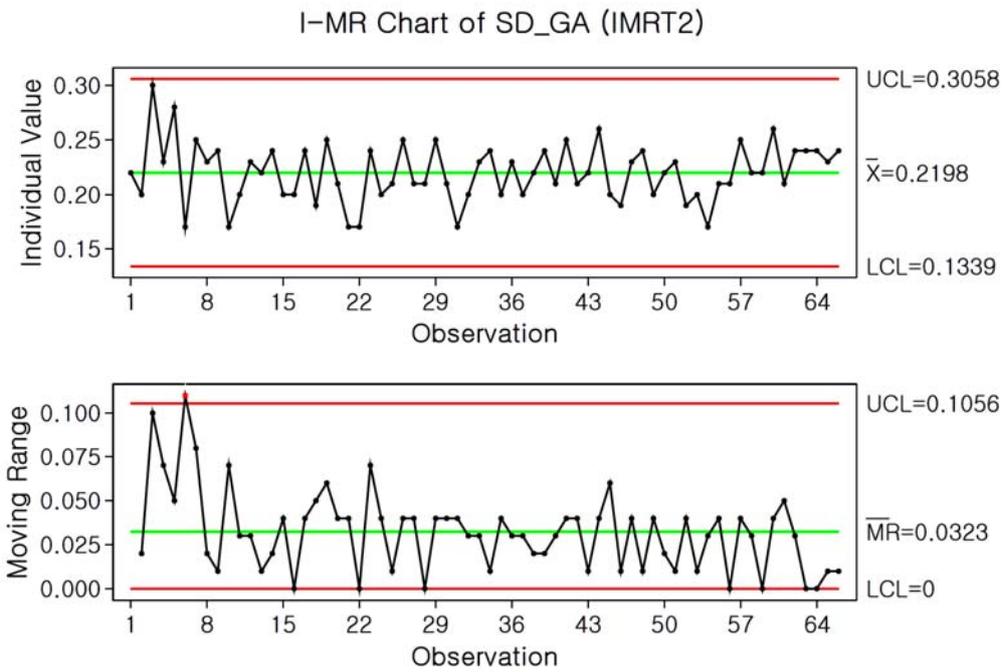

(b)

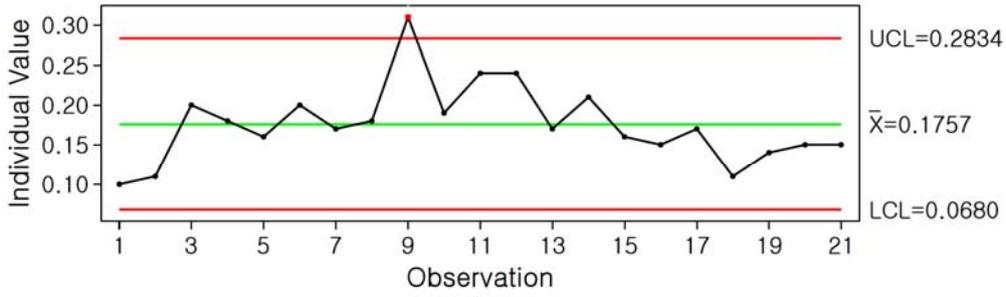

(c)

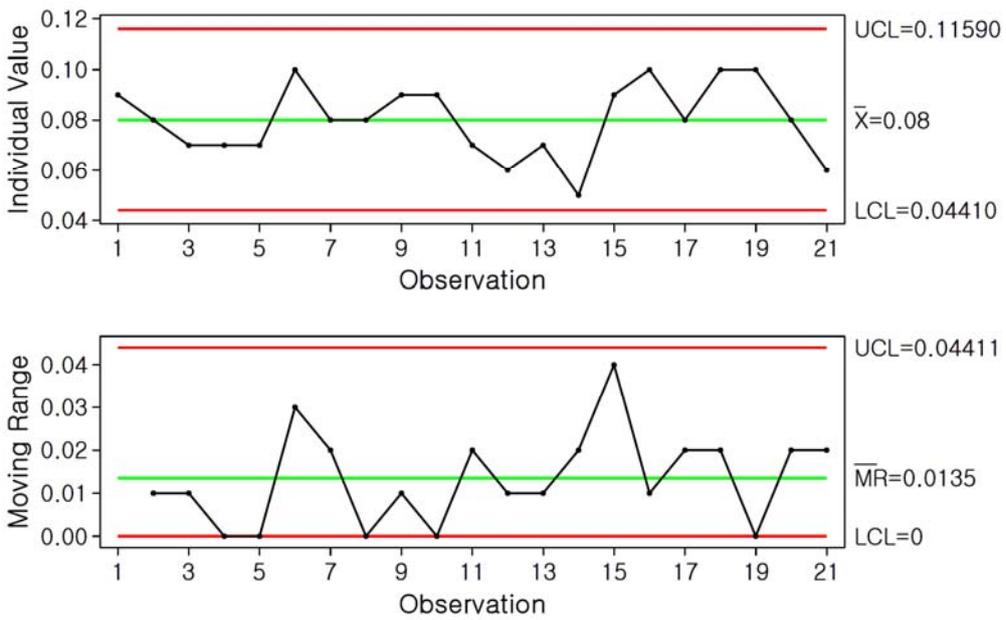

(d)

Figure 8. Capability histograms and distributions of σ$_{GA}$ for each group. The lower and upper specification levels (LSL and USL) were set to 0 and 0.5, respectively, for all cases.

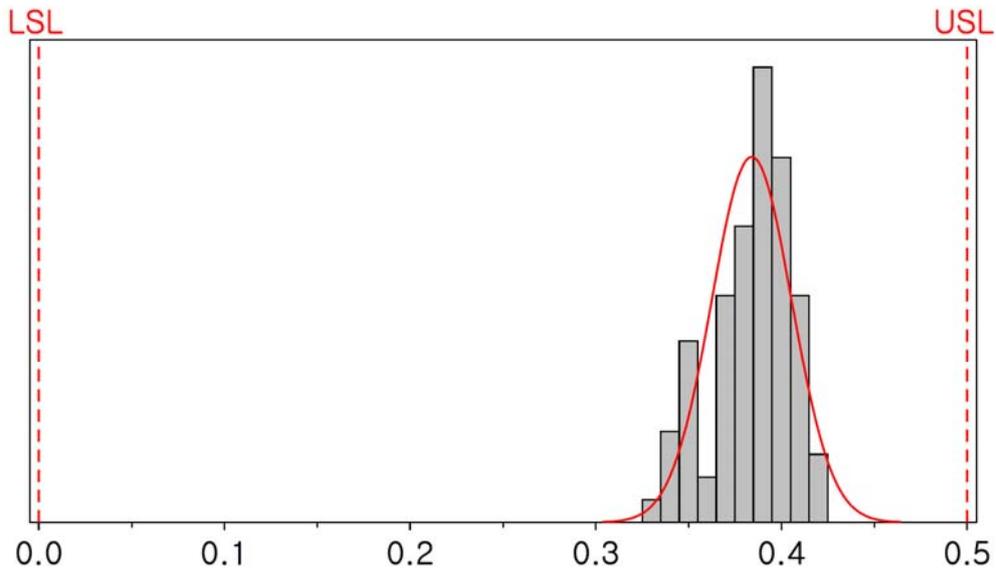

(a)

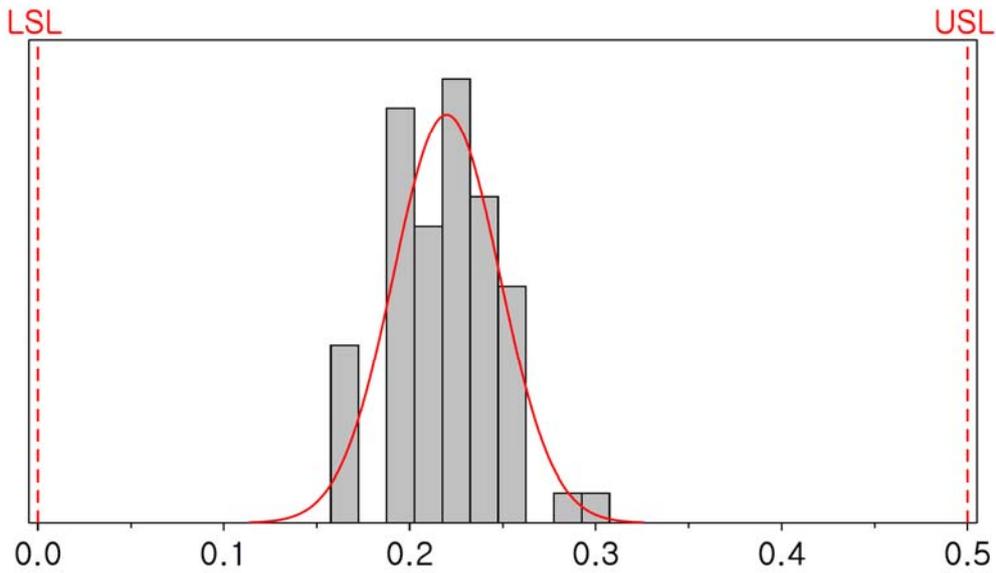

(b)

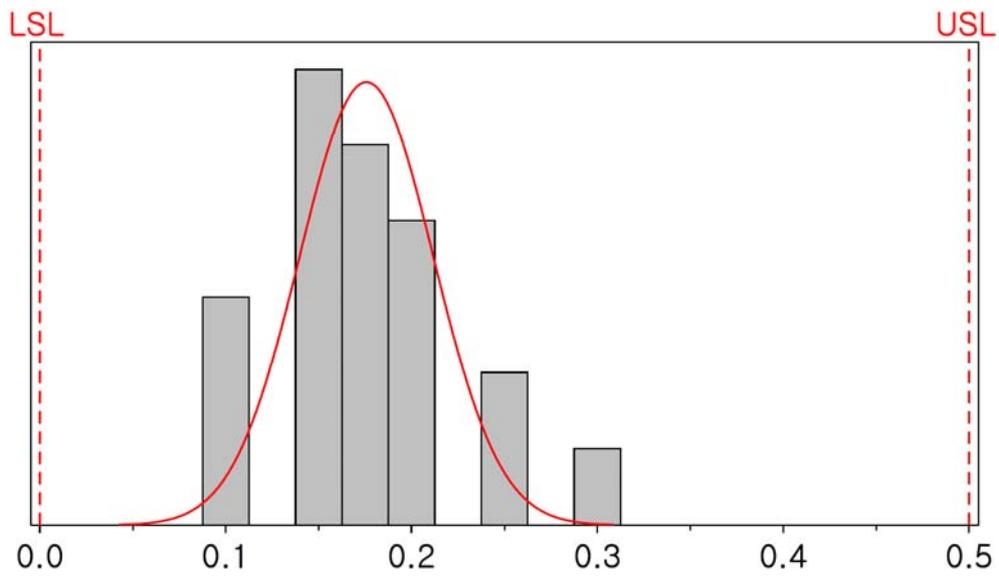

(c)

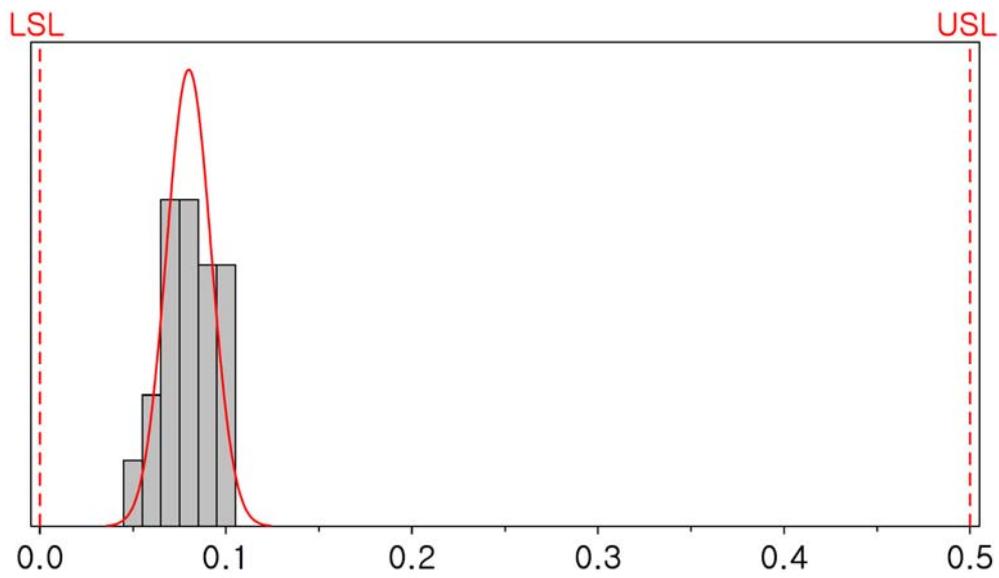

(d)